# Bond Synergy Model for Bond Energies in Alloy Oxides

Szu-Chia Chien and Wolfgang Windl

*Department of Materials Science and Engineering, The Ohio State University, Columbus, OH 43210*

**Abstract**

In this work we introduce a metal-oxide bond-energy model for alloy oxides based on pure-phase bond energies and bond synergy factors that describe the effect of alloying on the bond energy between cations and oxygen, an important quantity to understand formation and stability of passive films. This model is parameterized for binary cation-alloy oxides using density-functional theory energies and is shown to be directly transferable to multi-component alloy oxides. We parameterized the model for alloy oxide energies with metal cations that form the basis of corrosion resistant alloys, including Fe, Ni, Cr, Mo, Mn, W, Co, and Ru. We find that isoelectronic solutes allow quantification of pure-phase bond energies in oxides and that the calculated bond energy values give sensible results compared to common experience, including the role of Cr as the passive-layer former in Fe-Ni-Cr alloys for corrosion applications. Additionally, the bond synergy factors give insights into the mutual strengthening and weakening effects of alloying on cation-oxygen bonds and can be related to enthalpy of mixing and charge neutrality constraints. We demonstrate how charge neutrality can be identified and achieved by the oxidation states that the different cations assume depending on alloy composition and the presence of defects.

**Introduction**

High-entropy alloys (HEAs)[1,2], a new class of alloys that consist of five or more principal elements with nearly equiatomic ratios, have attracted significant interest because of their unique physical and chemical properties. HEAs were found to have high strength, good corrosion resistance and excellent mechanical properties at finite temperatures depending on their composition.[3,4] The high configurational entropy of random mixing was initially proposed to be a key to the stability of HEAs.[2] By now it has been found that this is probably not the case and many of the alloys do not represent their thermodynamic ground state, but are rather kinetically inhibited from phase decomposition and do so upon annealing.[1,5,6] Thus, in this paper, we understand the term HEA not just as entropy-maximizing alloys with equiatomic composition, but rather in the modern sense of multi-principal-element alloys (MPEA) where concentrations do not need to be equiatomic[3,7–11]. Their oxide equivalents, either high-entropy ceramics[12] or perovskite oxides[13], are also a currently very actively researched field. When an MPEA is exposed to an aqueous environment, a thin complex alloy oxide film forms on its surface due to strong thermodynamic driving forces.[14,15] This oxidation layer, also called a passive film, protects the alloy and is responsible for the corrosion resistance of corrosion resistant alloys. The oxidation process includes bond breaking in the metal and bond forming with oxygen at the surface or interface. Although the final oxide composition and structure is strongly influenced by a variety of factors, the bond strength between different metal and oxygen atoms plays an important role during the formation of the passive layer as a measure of the thermodynamic driving force. One method to estimate the metal-oxygen bond strength is to calculate it from the measured heat of sublimation.[16,17] However, while such values are available for single-cation oxides, it is not clear if they are transferable to alloy oxides and how they depend on the concentration of other cations. Because of that, we aim here to calculate the energetics in high-entropy alloy oxides through their bond energies, specifically the metal-oxygen bond strengths as well as the synergistic effects between different metal-oxygen bonds. In this work we propose a bond energy model as a basis for further study of the corrosion process and as a useful tool for designing corrosion resistant MPEAs. In order to understand the oxide formation better from a bond-energy point of view, it is necessary to address not only the oxide bonds, but also the metal-metal bonds in the alloys. An approach for calculating the latter for complex alloys was recently presented in the form of a bond-order bond energy model, where the alloy energetics are deconstructed into pairwise metal-metal bonds between all the elements in the alloy in the spirit of the quasichemical model[18]. Using density-function theory (DFT) alloy energies as input, different structures on the two ends of the pairwise phase diagrams are treated by separate bond energies and spliced together in the middle by a logistic-like function. This type of model was shown to describe the DFT energetics very well and the bond energies were found to be surprisingly concentration-independent.[18]

In this work we now develop an approach to determine metal-oxygen bond strengths for alloy oxides with mixed binary cation concentrations and examine its transferability to multi-component alloy oxides. As mentioned, the heat of sublimation is a related parameter.[16] However, during sublimation at a surface, the O atom is not fully coordinated, and so the bond energies are different from bulk bond energies, which are important to assess the stability of the passive film once formed. In principle, both cases should be studied. Figure 1 shows the relationship between calculated adhesion energies (where the minimum-energy position for the O atom is identified on the surface by trial and error and the energy difference between its relaxed surface position and far-away vacuum limit determines the adhesion energy) and metal-oxygen bond energies as determined later in this paper. With the exception of Mo, these two parameters scale closely with each other. This allows to use the easier-to-determine bulk bond energies to assess adhesion energies, the exact determination of which is computationally tedious.[17]

Modeling based on modern quantum-mechanics techniques such as DFT is an obvious approach to determine bond energies in alloy oxides, but the analysis of the results is not straightforward. DFT calculates total energies for the entire system, but usually does not easily allow identification of the energy associated with a particular bond. In this paper, we therefore investigate a holistic bond-energy model for metal-oxygen bond strength that can be fitted from total-energy calculations for a series of alloy oxides composed of the same elements, but with different composition. We will apply it to alloy oxides mostly involving chemical elements prevalent in corrosion resistant alloys. XPS experiments on the passive layer of 316 stainless steel show that the passive film formed within the first hour of passivation time contains Cr and Fe as the major cations, starting out at a ratio of ~1:1, increasing over time to Cr enrichment with a ratio of ~2:1, and with a cation/oxygen ratio on the surface of ~0.4, suggesting a surface stoichiometry of (Fe,Cr)$_2$O$_3$[19]. Since at that stoichiometry, corundum is the stable crystal structure for both Cr$_2$O$_3$[20] (which is the only stable room-temperature oxide of the Cr-O phase diagram) and Fe$_2$O$_3$[21], we will focus our effort on the corundum structure, which we find to be surprisingly stable for nearly all alloying elements. For comparison, we will also discuss results for rock salt, which is on the opposite end of the spectrum and only stable for NiO, MnO and CoO, but deteriorates for additions of small concentrations of solutes. Similar calculations can be made for other oxide structures, including those with amorphous features which have been observed in higher-Cr concentrations passive layers on FeCr alloys.[22] Such structures will be the focus of future work. In addition to Fe and Cr, we also include other common elements in corrosion resistant alloys: Co, Mo, Mn, Ni, Ru, and W, where Ru and W additions have been recently shown to result in especially corrosion resistant MPEAs.[3,8] We will also investigate the effects of charge compensation or charge neutrality in oxide alloys, showing first that alloys with combinations of cations having different oxidation states can result in overall charge neutrality with an energetic sweet-spot at the charge-neutrality composition. The cation ratio then allows determination of the relative oxidation states and helps to relate them to the Bader charges,[23,24] which can be calculated using DFT. Secondly, we examine in a similar way the effect of charge compensating vacancies and show how they can provide charge neutrality and stabilize the alloy especially when it contains charge-misfit cations.

**Bond Synergy Model**

When metal alloys oxidize, the different electronegativities and preferred oxidation states of the elements cause charge transfer between the different cations, changing the bond strength and thus bond energy between the cations and the oxygen anions. Therefore, to describe the bond energies in an alloy with an analytical model, a parameter is needed that indicates the effect of alloying on the different bonds, which we call bond synergy factor, $B$. It has been previously suggested that charge transfer in alloy oxides can be well described by a linear dependence[25], which we confirm for the examples of (Fe,Cr)$_2$O$_3$ and (Fe,Ni)$_2$O$_3$ alloys (see Figure 2 in the Results section). Since all metal oxides have fractional ionicity and thus have bonding intermediate between covalent and ionic types,[26] the bond strength dependence on charge should be between linear and quadratic.[27] Our results show that the linear dependence works sufficiently well, and we propose the following equation to fit the bond energy for alloy oxides,

$$E_{\text{tot}}(\text{M}_1, \ldots, \text{M}_N\text{-O}) = \sum_q \sum_{i=1}^N N_i^{(q)} \left[ \epsilon_{\text{M}_i\text{-O}}^{(q)} + \frac{1}{2} \sum_{j \neq i}^N x_j^{(q')} B_{ij}^{(qq')} \right], \quad (1)$$

where $q$ enumerates the different oxide bond types, differentiated by oxidation states or coordination (e.g. tetrahedral vs. octahedral); M$_1$, M$_2$, …, M$_N$ denote different metallic cations; O denotes oxygen, $N_i^{(q)}$ is the M$_i^{(q)}$-O bond count and $x_i^{(q)} = N_i^{(q)}/N_{\text{tot}}$ is the mole fraction of M$_i^{(q)}$; $\epsilon_{\text{M}_i\text{-O}}^{(q)}$ is the fitted pure-phase bond energy of the M$_i^{(q)}$-O bond; and $B_{ij}$ the bond synergy factor for the cation pair $M_i$ and $M_j$, which corrects the pure-bond energies in alloys due to changing charge transfer and bond lengths with composition. With that, a composition-corrected bond energy can be calculated,

$$E_{M_i-O}^{(q)}(\{x_j\}) = \frac{\partial E_{\text{tot}}}{\partial N_i^{(q)}} = \epsilon_{M_i-O}^{(q)} + \frac{1}{2}\sum_{j\neq i}^{N} x_j^{(q')} B_{ij}^{(qq')}. \qquad (2)$$

Since in corundum and rocksalt, all bonds are of the same type in terms of coordination and basic structure, we will drop the index $q$ in the remainder of the paper. In Eq. (1), the bond synergy factors, $B_{ij}$, are defined as absolute change (units of eV) rather than relative to the bond energies, which makes more sense for our purpose of examining mutual strengthening or weakening effects as we will discuss in the following. As a final not, for structures with different bond types $q$, additional strategies are needed to uniquely separate their energies from each other, such as adding other structures with only one bond type into the fitting base.

**Computational Methods**

The energy of the alloy oxides was calculated using DFT as implemented in the Vienna *Ab-initio* Simulation Package (VASP)[28,29] with projector augmented wave Perdew-Burke-Ernzerhof (PAW-PBE) potentials (for O, the soft variant was tested and then used for all calculations).[30] The DFT + Hubbard $U$ method[31] is used in this work, which combines DFT with a Hubbard-like term for the on-site Coulomb interactions of localized electrons. The $U$ parameters were taken from optimization efforts in previous publications and are used for the Fe,[32,33] Ni,[34] Cr,[35] W,[36] Ru,[37–40] Mn,[41] Co[42] and Mo[43] cations with values of 3, 5.3, 3.5, 2.2, 2, 4, 4.1 and 8.6 eV, respectively. The $U$ parameter of 2.2 for W was optimized to that value also under consideration of calculational convergence. All cells were fully relaxed with respect to ionic positions and lattice vectors, and an energy cut-off of 330 eV was employed in all calculations. For the cation distributions in the oxides, special quasirandom structures (SQS)[44,45] were prepared. For corundum, cells with 12 cations were used, while for rock salt, the cells had 32 cations. All calculations were performed within collinear magnetism, and for the initial configurations, the lowest energies among several antiferromagnetic configurations as well as the ferromagnetic configuration were determined. The lowest energy magnetic configurations of most binary cation oxides studied in this work had antiferromagnetic spin structures, which includes most pure metal oxides such as $Fe_2O_3$, $Cr_2O_3$, $Ni_2O_3$, $W_2O_3$ and $Mo_2O_3$. $Co_2O_3$ and $Ru_2O_3$ are found non-magnetic, while $Mn_2O_3$ has ferromagnetic spin ordering, although in the corundum structure which is not one of its ground state phases. Some binary oxides were found to have intermediate non-zero magnetization, especially for high-Ni and high-Co compositions. A 3×3×2 Γ-centered (2×2×2 Monkhorst-Pack[46]) k-point mesh was used for corundum (rock salt). From the total energies, the energies of all constituent atoms in vacuum were subtracted, also calculated with VASP for 20 Å × 20 Å × 20 Å boxes. After that subtraction, an oxide cell with infinite lattice constant would result in zero total energy, which means that the resulting numbers represent the sum of the bond energies.

We also calculated the energy of alloys in multi-component systems to evaluate whether the bond energies from the binary-oxide fits can extend to multi-component alloys in the spirit of an interatomic pair potential. For this test, larger corundum structures consisting of 24 metal cations and 36 oxygen atoms were prepared with different metal compositions. The sum of the bond energies was then calculated using DFT calculations from the total energy in the same way and compared to the values from the bond energies fitted for the cation-binaries. The same settings for the DFT calculations were used as for the smaller cells, except for a 3×2×1 Γ-centered k-point mesh.

**Crystal Structures Considered**

In this work, we first applied the present bond synergy model to one of the most common structures in passive films, corundum. Corundum, shown in Figure 3(a), has a hexagonal (rhombohedral) crystal structure with $M_2O_3$ stoichiometry, and is found for crystals such as $Al_2O_3$, $Cr_2O_3$, and $Fe_2O_3$.[47,48] The unit cell contains six formula units. In corundum, the oxygen atoms, nominally $O^{2-}$, form an hcp lattice, where two thirds of the octahedral voids are occupied by the cations, nominally $M^{3+}$. Each cation in the corundum structure is surrounded by 6 oxygen atoms, resulting in coordination numbers of 6 for cations and 4 for anions. All alloy structures considered were found to be stable and maintained their corundum structure upon relaxation with small deviations caused by the random cation distribution.

Next, we applied this model to the rock salt (or NaCl) structure, shown in Figure 3(b). It is composed of two ions located on fcc lattices that are shifted by (0.5, 0.5, 0.5) from each other. Both ions have six neighbors in octahedral coordination. Whereas all corundum structures studied were well-behaved in our calculations and allowed high-quality fitting with Eq. (1), the situation is different for the rock salt structure. While NiO and CoO are both line compounds, MnO has a wider solubility range on the oxygen-rich side where the composition ranges from MnO to $MnO_{1.045}$[49]. Because Ni is commonly present at higher concentrations in corrosion resistant alloys as compared to Co and Mn, we chose NiO as our base solvent with solutes in relatively small concentrations. Specifically, we found that none of the solutes destabilizes the NiO rocksalt structure in the range between 3.125% and 12.5% (i.e., 1-4 solute atoms in 32 cation rock salt cells) and thus restricted our calculations for all solutes to that range.

**Results**

In order to validate our choice of DFT parameters, especially the values for the Hubbard U's, we first examine the oxide phase stability in the oxygen concentration range between ½ and 2/3 for rock salt (MO), corundum ($M_2O_3$), spinel ($M_3O_4$) and rutile ($MO_2$) structures. For that, we calculate the convex hulls for the different elemental oxides. In a convex hull construction between elements A and B, the formation energy per atom, $E_{form}(A_nB_m) = [E_{tot}(A_nB_m) - nE_{at}(A_{bulk}) - mE_{at}(B_{bulk})]/(n + m)$, is plotted as a function of composition. The lowest-energy points for each composition are then connected in a convex hull[50,51], which is the shape when a rubber band held the end points is first pulled down below all points and then released towards the points. The points that end up on the convex hull then indicate stable phases, while points above it are unstable.

Figure 4 shows the convex hulls for our set of oxides and the metal cations considered in this work, based on zero-temperature, zero-pressure total energy calculations without entropic contributions. With two exceptions, all phases are accurately predicted in agreement with the different phase diagrams[49] by our calculations. The two exceptions are FeO in the rocksalt structure and $Mn_2O_3$ in the corundum structure, both of which do not appear in their respective phase diagrams. Their energies per atom are 27 and 35 meV below the convex hulls formed by their neighbors, respectively, which gives us an estimate of our numerical accuracy which is compatible with the typical errors of ~0.01 eV per atom, provided the discrepancy is not a consequence of the neglected vibrational entropy. Since we adopted the settings for iron oxides from Kresse and Joubert[30], which is one of the most thorough and exhaustive investigations of the effect of DFT settings on predicted energies, this may be possible. Overall, the otherwise correctly predicted phases for a total of eight oxide systems and four crystal structures can be taken as strong validation of our DFT methodology and gives us confidence in a sensible prediction of the examined energy values.

For corundum, we find that the alloy energetics can be excellently fitted with Eq. (1). Results for the fitted average bond energies are shown in Figure S1, while pure-phase bond energies and bond-synergy factors are shown in Figure 5 (the tabulated values can be found in Tables 1 and 2). Among all pure-phase bond energies (Figure 5(a)), Ni-O is the weakest, whereas W-O is the strongest and Cr-O is the second strongest. It may be tempting thus to think about tungsten oxide with a corundum structure as a highly protective passive layer. However tungsten oxide does not naturally occur in the corundum structure due to its preference for high oxidation states. Still, the experimentally observed elevated W-content in Cr-rich oxide layers in comparison to that in the bulk alloy[8] corroborates our finding of strong W-O bonding. Note that since cobalt has only few highly specialized uses in alloy steels, we restricted our study here on its interaction with the other elements found in Cantor MPEAs[52], which also contain Ni, Fe, Cr and Mn.

Ni and Cr are the most common alloying elements in stainless steels (in addition to Fe). Among these three elements, Cr shows the highest pure-phase bond energy in corundum. At the same time, the bonds that Cr forms in the metal alloy with Cr, Ni, and Fe are ~0.1 eV weaker than Fe and Ni bonds[18], thus making Cr the most favorable element to switch from metal alloy to oxide. This is consistent with the observation that a significantly enriched Cr concentration is usually found in the passive films on both FeCr and FeNiCr alloys[53,54] and is at the heart of the excellent corrosion resistance in such alloys. Furthermore, no indication of $Ni^{3+}$ with corundum structure is found in experiments except for systems far away from thermodynamic equilibrium[55]. While it is observed that, with increasing Cr content, the passive layer changes from crystalline to amorphous structure in FeCr alloys,[22] we do not consider this phase change in the present study.

Whereas this argument easily explains the abundance of Cr in the passive layer, it cannot be extended in a straightforward way to W, although its bonds are 20% stronger than for Cr. Even though the W-O bonds are more favorable than Cr, Fe, and Ni bonds by 0.4-0.8 eV, the bonds of W to Fe, Ni, and Cr in the metal alloy are also stronger by ~0.4 eV.[38] Thus, an explicit overall energy balance for a specific alloy is necessary to decide the thermodynamically favorable oxide composition.

The bond-synergy factors, which are shown in Figure 5(b), can be divided into two types, the *bond weakeners* (top) and the *bond strengtheners* (bottom). Bond strengtheners increase the absolute values of the pure-phase bond energies, while bond weakeners have the opposite effect and destabilize the oxide.

Bond weakening and strengthening is directly related to the enthalpy of mixing, as a comparison between Figures S1 and S2 shows. A positive enthalpy of mixing indicates miscibility problems between the constituents (unless overcome by sufficient entropy from vibrations and mixing[9]) and phase separation, which is equivalent to bond weakening. Even though often small, bond weakening is found over the entire composition range for 15 binary systems, Mn-Mo, Mn-Fe, Fe-Mo, Ru-Mo, Mo-Cr, Mn-Cr, Mo-Ni, W-Mo, Fe-Ni, Fe-Cr, Co-Cr, Co-Fe, Ru-Fe, W-Cr, and Ru-Cr. This list includes all systems with Cr. Thus, in thermodynamic equilibrium, Cr has the tendency to dominate as a component in oxides with corundum structure, as is well known from experiment.[56] On the other hand, systems with bond strengthening display a negative enthalpy of mixing. While Co-Mn and Ni-Cr have small mixing enthalpies of just a few hundredths of an eV, which is close to the DFT error and makes these systems somewhat hard to analyze, the enthalpy of mixing curves for W-Ni, Mn-W, Ru-Ni, W-Fe, Ru-Mn, and Mn-Ni have a distinct "V" shape for at least for a part of the composition range, which as we will show can be understood by the preferred oxidation states of the cations and charge neutrality.

In order to understand the role of charge neutrality in the alloy oxides under consideration, we have to look at the oxidation states of the considered structures. In corundum, the stable oxidation state of O is –2, and the cation oxidation state is +3.[57] For oxides in the rock salt structure, the values are –2 and +2, respectively[58]. In a charge-neutral alloy oxide, the cations do not necessarily all need to have oxidation states of +3 or +2, respectively, but need to average out to these values. In order to estimate the oxidation states from DFT, a number of approaches exist[58]. We found the recent suggestion[59] of determining oxidation states from Bader charges, where the mapping has been derived from machine learning, most practical for the alloys considered here. Bader charges of atoms in a solid are calculated by integrating the electron density within their surrounding volumes, which are limited by the 2-D surface on which the charge density has a minimum perpendicular to the surface[23,24]. The results from the literature[59] are compared in Table 3 to the results from the present work where the oxidation states could be identified as described below. Oxygen with a Bader charge of < –0.96 electron charges has oxidation state –2, while it is -1 or 0 for larger values[59]. With our DFT settings, it seems that the threshold to oxidation state –2 is similar, around -1.0, as we will discuss in the following.

For a number of corundum systems, both cations are clearly in the +3 oxidation states, while oxygen is –2. In these systems, both cations and oxygen maintain with small error the same Bader charge within or close to the range identified previously[59], while the oxygen charge is between −1.2 and −1.1 and only shows small variations. These systems include Fe-Cr, Mo-Cr, Fe-Mo, W-Mo, and Mn-Fe (All Bader charge values are listed in Table 4).

We consider next the Fe-Ni, Ni-Cr, Ru-Cr and Ru-Mo systems where both cations keep their charges throughout the entire composition range, but only one of them is +3, and O changes its Bader charge nearly perfectly linearly from a value below the -2 threshold to one that is above it by more than 0.2 (Table 5). An example for this is Ru-Cr. Over the composition range, Ru and Cr maintain Bader charges of 1.42± 0.01 and 1.72 ± 0.003, respectively. Since for $Cr_2O_3$, the oxygen charge is –1.16, while it is –0.94 for $Ru_2O_3$, we can attribute oxidation states of +3 for Cr and +2 for Ru. Since no charge compensation can happen in this system, the enthalpy of mixing is small. Another system of this type is Ni-Fe, where the Ni charge is 1.22-1.27 and the Fe charge is 1.66-1.69, while the O charge changes from −0.85 on the Ni end to −1.11 on the Fe end. Similarly, in Ni-Cr, the Ni charge is 1.22-1.27 and the Cr charge is 1.74-1.84, while the O charge changes from −0.85 on the Ni end to −1.16 on the Cr end. From the viewpoint of charge compensation, these systems may only be realized in the presence of charge-neutralizing defects, which will be briefly discussed below. A similar situation exists for cases where both cations requires oxidation states larger than +3, especially Ru-Mo where no charge compensation is possible and a nearly ideal mixing enthalpy curve is found. As a consequence, all bond-weakening enthalpies have maxima smaller than 0.25 eV.

Finally, there are systems with charge compensation for specific compositions, resulting in a distinct minimum in the enthalpy of mixing and thus bond strengthening. An example for this is W-Ni. The Bader charge on W for 16.7% and 25% W is 2.64 and 2.63, respectively. For higher concentrations it drops continuously to a value of 1.61 for pure $W_2O_3$. Simultaneously, the charge on Ni changes little with a value of 1.18 ± 0.04, which again puts it in its preferred oxidation state of +2, while the charge on O drops initially sharply with W content and crosses the threshold for oxidation state -2 at a W concentration of 25% with a value of -1.04. Thus at 25% W, W is still in its preferred oxidation state of +6 as characterized by its Bader charge in reference to Table 3 and by considering that Ni and O are in oxidation states +2 and -2, respectively, which results in neutral formal charge at 25% W. For higher W content, the W charge approaches the value of 1.61 corresponding to oxidation state +3, which is not one of the preferred states of W, which then results in an energy penalty. As a result of these effects, the formation enthalpy has a V-shape with a distinct minimum at 25% W. Similarly, Mn-W also has a concave-shape in the enthalpy of mixing with its minimum also at 25% W. Bader charges on W for 16.7% and 25% W are 2.66 and 2.63, respectively. Simultaneously, the charge on Mn changes little with a value of 1.50 ± 0.03, which again falls in its most common oxidation state of +2. On the other hand, the W-Fe system has a minimum at 33% W. While Fe has a Bader charge of 1.37 and thus a +2 oxidation state, the Bader charge of W is 2.33, between the values of +4 (2.14-2.27) and +6 (2.69-3.04). This suggests that W is in the somewhat unusual oxidation state of +5, which indeed agrees with the minimum in the mixing enthalpy at 33%, since 2/3×(+2) + 1/3×(+5) is close to +3. These results suggest that "magical" alloy compensations exist in these systems that are fully charge compensated, which are yet to be confirmed experimentally.

An interesting case in this category is Ru-Ni, which also has a V-shape in its enthalpy of mixing, though there is a broad minimum shared between 33% and 50% Ru. The Bader charge on Ni is 1.21-1.22, suggesting the oxidation state of Ni is +2. However, the Bader charge on Ru changes from 1.89 at 33% Ru to 1.69 at 50% Ru. For both compositions, the Bader charge for O is below the -0.96 threshold for the -2-oxidation state. According to Table 3, the Bader charge of 1.69 corresponds to the +4-oxidation state, which is sensible for the 50% charge-compensated alloy. However, the value of 1.89 is between the tabulated values of +4 and +6, suggesting an oxidation state of +5, which is, as previously discussed, expected for a minimum at 33% Ru. Thus, it seems that Ru can adopt both +4 and +5 oxidation states when alloyed in corundum with Ni, allowing for charge-compensated compositions in the 33%-50% range. A similar analysis can be made for Ru-Mn. In all the V-shaped systems, the extremal enthalpy of mixing for this system is large, between ~ –0.2 and –1.3 eV.

Other than cation combinations of species whose oxidation average to +3, another possibility of charge compensation, which was just shown to be a major driving factor in alloy oxides, is the presence of vacancies. As an example, we have performed calculations of bond energies in the Fe-Cr-Ni system with oxygen-vacancies, where the fit was performed by including explicit bonds between the cations and the vacancy. While the change in M-O bond energies for alloys that contain only Fe and/or Cr is very small (on the order of 0.3%-0.5%), the presence of vacancies has a large effect on Ni-O bonds. Our results concur with the usual assumption of one O vacancy providing two extra positive charges and thus allowing Ni to go from the +2 oxidation state to +3 to minimize the Ni-O bond energies in the system. With a ratio of 2 between the number of Ni cations and the number of O vacancies, the Ni-O bond energy changes from -1.7 eV to -2.0 eV, a value that is in agreement with the W-Ni defect-free system, where the bond-synergy factor changes the Ni-O bond energy from -1.7 to -2.0 eV for the charge-compensating composition of 1/4 W. Also, the corresponding panel in Figure S2 shows that the Cr-Ni formation energy curve changes from positive values close to zero (no vacancy) to a charge-compensated V-shape with a strong minimum of -2.3 eV for 2 Ni atoms (one O-vacancy). The Bader charges further support this oxidation state picture: while a vacancy of pure O changes the average charge on Cr from the perfect value of 1.75 to 1.66 and maintains the ideal O-charge of -1.17, the Cr-charge is restored to 1.74 in the presence of a vacancy and 2 Ni cations while the O charge is maintained at a value of -1.15. If two more Ni-ions are added, both Cr and O charges deviate from the ideal level to values of 1.82 and -1.08 respectively, and the situation gets worse with increasing Ni concentration. Looking at the overall picture of the vacancy case in combination with our previous results, we find that, independently of the method of charge compensation, all cation-oxygen bonds in charge-compensated Fe-Cr-Ni alloy oxides have bond energies of ~-2.0 eV. The presence of O-vacancies in Ni-doped corundum is in agreement with experimental observations[57].

While these examples show that the bond-synergy model gives interesting results for the interplay between different cations in binary alloy oxides, it would be even more valuable if it were also transferable to multicomponent alloys, which would mean that the alloy energetics can be decomposed in pairwise terms and thus could be well described by the bond energies and bond synergy factors developed for binary alloy oxides. In order to examine if this is indeed the case, we tested how the bond synergy model with parameters fitted to binary oxides performs for 18 multi-component alloy oxides containing between three and six components, whose compositions were chosen randomly to cover a sensible range (compositions listed in Table 6). Figure 6 shows the comparison of average bond energies obtained from DFT and the predicted average bond energies from the bond synergy model with parameters from the binary fits. We find excellent agreement between DFT and bond synergy model results, strongly suggesting that the bond energies and bond synergy factors obtained from binary systems are fully transferable to multi-component alloy oxides.

After studying corundum, which we found to be an extremely accommodating structure for alloying, we test the bond synergy model now for rock salt, a much less inclusive crystal structure. As rationalized in the section on crystal structures, we here discuss results for Ni-based binary alloys in the rock salt structure with small solute concentrations up to 12.5%. For these oxides, once again the alloy energies can be fitted very well with Eq. (1). The fitted bond energies and bond synergy factors are shown in Figure 7 and the corresponding values in Table 7 in the first two columns.

Considering the absolute values, we find that Mo-O bonds have the strongest energy in the rock salt structure. In the considered composition range, only the Ni-alloy with Co shows a weak bond strengthening effect (bond synergy factor ~ -0.22 eV), whereas all other cation-Ni combinations exhibit bond-weakening effects. In order to understand these effects, we once again look at the Bader charges. Comparing the charges in the pure NiO phase with known oxidation states of +2 for Ni and -2 for O, we find that a Bader charge of 1.29 for Ni and -1.29 for O in this structure corresponds to oxidation states of +2 and -2. For the investigated concentrations up to 12.5% of Fe, Cr, Mn, Ru and Co, we find that the Bader charges on Ni and O both stay nearly constant with values of 1.28-1.29 for Ni and -1.30 to -1.28 for O with standard deviations of 0.01 or smaller. Following our discussion for corundum, this strongly suggests that the impurity cations are also assuming a +2 oxidation state over the entire concentration range and puts the Bader charges indicating that oxidation state at 1.30±0.05, 1.48±0.03, 1.33±0.02, 1.34±0.09 and 1.38±0.03 for Fe, Cr, Mn, Ru and Co, respectively, as summarized in Table 3.

On the other hand, the two classic refractory metals Mo and W assume higher oxidations states with charges of ~1.8. A comparison with Table 3 suggests they are in oxidation state +3. When their concentration is higher, their high charge forces the amount of charge on Ni and O to decrease (see Table 8), for the 12.5% solute cases, to $Q_{Ni}$ = 1.14 in both cases and $Q_O$ = -1.20 and -1.22 for W and Mo solutes, respectively, destabilizing the structure.

The bond weakening effect for all solutes except Co is not surprising. Co has rock salt (albeit as a high-temperature phase) in its phase diagram with O and, while smaller than Ni, has similar elastic properties within 15%. MnO exists also in the rock salt structure, but its large lattice constant (6.5% larger than NiO) in combination with its high compressibility (43% larger than NiO) result in a bond synergy on the bond-weakening side. All other solutes do not occur naturally in the rock salt structure and their presence thus destabilizes NiO. This suggests that NiO passive films should be most stable without solutes. Those should then phase separate into other structures. In turn, alloyed rock salt films that form due to kinetics should decrease the energetic gain for Ni to form into this crystal structure. We will show in the next section that this indeed can be observed in experiments.

**Bond Synergy Effects in Experiments**

While the non-linearities observed in the concentration-dependent energies in Figures S1 and S3 can be taken as an indication that a non-linear bond energy method should be more sensible, one could still argue that the non-linearities look small, and a linear fit may describe the overall energies at a simpler but still satisfactory level. To show that this is not the case and that the results from the bond synergy model are in agreement with experimental observation while those from a linear model are not, we also performed a linear fit of the bond energies without considering bond synergy factors for rock salt NiO, where the non-linearities are significantly more pronounced than in the corundum structure. The resulting values are shown in the last column of Table 7, along with a bar chart in Figure 8. Compared to the bond-synergy derived values, the strength of the metal-oxygen bonds changes considerably. At the extreme ends, without bond synergy, the pure-phase bonds in Mo are weaker by a factor of 2, while for Co, the bonds appear 25% stronger. Besides the quantitative changes, also the order of the bond energies changes dramatically, which allows us to validate our bond-synergy model in comparison to experimental findings, for which the two bond weakeners, Mo and W, are the most suitable candidates. While W-O stays stronger than Ni-O independent of consideration of bond synergies, Mo moves from the strongest non-Ni-O bond (with bond synergy) to weakest bond (without). If the model without bond synergy were valid, W should strongly compete with Ni in rock salt oxides and decrease the fraction of $Ni^{2+}$ cations, while Mo should have no or negligible effects. By considering bond-synergy, Mo is also a bond weakener and its oxide-bond energy is 0.5 eV stronger than Ni-O.

Comparing these expectations with experiments, the effect of bond synergy can be seen in NiCrMo alloys, where both rock salt (NiO) and corundum ($Cr_2O_3$) oxides are formed.[55] It is found that both Mo and W alloying decreases the fraction of $Ni^{2+}$ in the oxides formed on NiCr based alloys[60] as measured by XPS. Our present results suggest two explanations for this. First, Mo and W both weaken the Ni-O bonds in rock salt and thus make that structure less favorable for Ni. At the same time, the bond weakening effects of Mo and W on Cr-O in corundum are much weaker and thus much less destabilizing compared to NiO. In addition, the strong bond strengthening effect between W and Ni in corundum makes corundum a much more attractive destination for Ni.

Considering these points, it is no surprise that W or Mo additions increase the measured fraction of $Cr^{3+}$ and decrease in $Ni^{2+}$.[60] In summary, these findings can be understood as a clear confirmation of the predicted effect of Mo and W on NiO and $Cr_2O_3$ oxide energetics within the bond synergy model.

**Conclusions**

In this paper we introduced a new metal-oxygen bond-energy model for alloy oxides based on pure-phase bond energies and bond synergy factors that describe the effect of alloying on the bond energy between cations and oxygen. We parameterized the model using density-functional theory results for multi-component alloy oxide energies. Two common crystal structures found in passive layers on alloys, corundum and rock salt, were studied with metal cations that form the basis of stainless steels (Fe, Ni, Cr) as well as other common elements in corrosion resistant alloys.

We found that the calculated bond energy values give sensible results in comparison to common experience, such as Cr being the passive-layer former in FeNiCr alloys. Also, the bond synergy factors suggest mutual strengthening and weakening effects of alloying on cation-oxygen bonds. This bond strengthening and weakening effects found in the bond-synergy factor are related to charge compensation, which is reflected in the enthalpy of mixing in binary alloy oxides. Besides alloys where both cations assume the alloy-optimal oxidation state (such as +3 in corundum and +2 in rocksalt structure), cations with different oxidation states can produce an energetic sweet-spot at the charge-neutrality composition, for example Ni-W in corundum. Another charge-compensating mechanism can be the presence of vacancies, which inject additional positive charges into the alloy and can compensate for cations whose original oxidation state is below the structure-optimal value. In the technologically most important Fe-Cr-Ni alloy system, we find a "universal" bond energy of ~-2.0 eV between cations and oxygen in the corundum system for fully charge-compensated systems, independent of the compensation mechanism. We also showed that O vacancies can be used as a probe to check the unknown cation oxidation states, as we showed for the Ni-Ru rocksalt system where Ru is surprisingly found in an oxidation state of +1 rather than +2.

For rock salt in the limit of Ni-rich oxide alloys, we found that Co-Ni is the only bond strengthener while all other cations weaken the Ni-O bond including Mo and W. This explains why in XPS measurements for oxides on NiCr alloys with additions of Mo and Ni, a decrease of $Ni^{2+}$ intensity is found. Overall, while the oxide composition in passive layers is of course strongly affected by kinetics and environmental conditions, we find that the bond synergy model and its derived energies are a simple and intuitive basis to examine fundamental thermodynamics in oxide films.

**Acknowledgments**

This work was funded by the Center for Performance and Design of Nuclear Waste Forms and Containers, an Energy Frontier Research Center funded by the U.S. Department of Energy, Office of Science, Basic Energy Sciences under Award # DESC0016584. The authors thank Ohio Supercomputer Center (OSC) for computational resources under project # PAA0010.

**Table 1**. Pure-phase bond energies for corundum alloy oxides calculated from a fit of Eq. (1) to the bond energies shown in Fig. 5.

| Metal-oxygen bond | Pure-phase bond energy (eV) |
|---|---|
| Ni-O | -1.6936 |
| Co-O | -1.7939 |
| Mn-O | -1.8824 |
| Mo-O | -1.9102 |
| Ru-O | -2.0128 |
| Fe-O | -2.0129 |
| Cr-O | -2.0838 |
| W-O | -2.4856 |

**Table 2**. Bond-synergy factors for corundum alloy oxides calculated from a fit of Eq. (1) to the bond energies shown in Figure 5.

| Cation Pair | Bond synergy factor (eV) | Cation Pair | Bond synergy factor (eV) |
|---|---|---|---|
| Ni-Fe | 0.0443 | Mn-W | -0.4285 |
| Ni-Cr | -0.0121 | Mn-Ni | -0.0922 |
| Fe-Cr | 0.0353 | Mn-Fe | 0.0951 |
| Mo-Cr | 0.0647 | W-Mo | 0.0466 |
| Mo-Ni | 0.0515 | W-Fe | -0.1686 |
| Mo-Fe | 0.0869 | W-Ni | -0.7124 |
| Ru-Mo | 0.0732 | W-Cr | 0.0073 |
| Ru-Fe | 0.0263 | W-Ru | -0.1383 |
| Ru-Ni | -0.3362 | Co-Fe | 0.0273 |
| Ru-Cr | 0.0022 | Co-Cr | 0.0307 |
| Mn-Ru | -0.0909 | Co-Ni | -0.0076 |
| Mn-Cr | 0.0566 | Co-Mn | -0.0226 |
| Mn-Mo | 0.1229 | | |

**Table 3.** Bader charges and their correspondence to the different oxidation states (OS) as determined in Ref. 51 and from the results in the present work where the oxidation state could be identified. All charges are in units of the elementary charge. For cations whose charges are different for the same oxidation states in the different crystal structures, "cor" denotes corundum and "RS" rocksalt structure.

| | From Ref. 51[59] | | | | This Work | | | |
|---|---|---|---|---|---|---|---|---|
| **Element** | OS +2 | OS +3 | OS +4 | OS +6 | OS +2 | OS +3 | OS +4 | OS +6 |
| Fe (cor) | 1.380 | 1.43-1.84 | | | 1.43 | 1.67-1.70 | | |
| Fe (RS) | | | | | 1.30±0.04 | | | |
| Ni (cor) | 1.220 | 1.33-1.39 | | | 1.21-1.27 | | | |
| Ni (RS) | | | | | 1.29±0.002 | | | |
| Cr (cor) | 1.340 | 1.76 | 1.83-1.94 | | 1.40 | 1.74-1.76 | | |
| Cr (RS) | | | | | 1.48±0.03 | | | |
| Mo | | | 2.08-2.24 | 2.58-2.68 | 1.43 | 1.70-1.81 | | |
| Mn (cor) | 1.36-1.56 | 1.68-1.73 | 1.83-1.91 | | 1.25 | 1.66-1.68 | | |
| Mn (RS) | | | | | 1.33±0.02 | | | |
| W | | | 2.14-2.27 | 2.69-3.04 | 1.39 | 1.62-1.75 | 2.13±0.02 | 2.64±0.01 |
| Ru (cor) | | | 1.65-1.85 | | 1.35 | 1.41-1.43 (O: -0.94) | | |
| Ru (RS) | | | | | 1.34±0.09 | | | |
| Co (cor) | 1.14-1.28 | | 1.43 | | 1.35-1.36 | 1.40-1.48 | | |
| Co (RS) | | | | | 1.38±0.03 | | | |
| All | 0.73-1.97 | 0.86-2.47 | 1.33-3.2 | 2.58-3.47 | | | | |

Table 4. Bader charges of binary alloy oxides where both cations are at +3 oxidation states.

| $Fe_x$ | $Cr_{1-x}$ | O |
|---|---|---|
| 1.698-0.032x | 1.745+0.014x | -1.167+0.051x |
| $Mo_x$ | $Cr_{1-x}$ | O |
| 1.787+0.014x | 1.746+0.021x | -1.164-0.037x |
| $Fe_x$ | $Mo_{1-x}$ | O |
| 1.710-0.041x | 1.804-0.017x | -1.204+0.090x |
| $W_x$ | $Mo_{1-x}$ | O |
| 1.800-0.134x | 1.814-0.106x | -1.215+0.107x |
| $Mn_x$ | $Fe_{1-x}$ | O |
| 1.673-0.00 x | 1.673-0.005x | -1.115+0.001x |

**Table 5.** Bader charges in non-compensating corundum alloy oxides with cations with oxidation states +3 and +2.

| $Fe_{1-x}$ | $Ni_x$ | O |
|---|---|---|
| 1.672+0.015$x$ | 1.192+0.073$x$ | -1.108+0.296$x$ |
| $Ni_x$ | $Cr_{1-x}$ | O |
| 1.207+0.058$x$ | 1.751+0.120$x$ | -1.171+0.320$x$ |
| $Ru_x$ | $Cr_{1-x}$ | O |
| 1.435-0.022$x$ | 1.744-0.007$x$ | -1.165+0.222$x$ |
| $Ru_x$ | $Mo_{1-x}$ | O |
| 1.415-0.007$x$ | 1.798+0.008$x$ | -1.201+0.262$x$ |

**Table 6**. Composition of cations shown in Figure 6.

| Test | Composition | | | | | | | |
|---|---|---|---|---|---|---|---|---|
| | **Cr** | **Ru** | **Mo** | **Fe** | **W** | **Ni** | **Mn** | **Co** |
| 1 | 11 | 7 | 2 | 2 | 1 | 1 | 0 | 0 |
| 2 | 11 | 7 | 1 | 2 | 2 | 1 | 0 | 0 |
| 3 | 11 | 2 | 2 | 7 | 1 | 1 | 0 | 0 |
| 4 | 2 | 7 | 2 | 11 | 1 | 1 | 0 | 0 |
| 5 | 2 | 2 | 7 | 11 | 1 | 1 | 0 | 0 |
| 6 | 11 | 2 | 7 | 2 | 1 | 1 | 0 | 0 |
| 7 | 4 | 0 | 0 | 9 | 0 | 11 | 0 | 0 |
| 8 | 11 | 0 | 0 | 9 | 0 | 4 | 0 | 0 |
| 9 | 8 | 0 | 8 | 0 | 0 | 8 | 0 | 0 |
| 10 | 11 | 0 | 0 | 9 | 4 | 0 | 0 | 0 |
| 11 | 7 | 3 | 0 | 14 | 0 | 0 | 0 | 0 |
| 12 | 6 | 0 | 4 | 14 | 0 | 0 | 0 | 0 |
| 13 | 6 | 0 | 5 | 0 | 0 | 13 | 0 | 0 |
| 14 | 6 | 0 | 0 | 0 | 0 | 13 | 5 | 0 |
| 15 | 6 | 0 | 0 | 0 | 0 | 15 | 3 | 0 |
| 16 | 6 | 0 | 0 | 0 | 0 | 10 | 8 | 0 |
| 17 | 6 | 0 | 0 | 0 | 0 | 5 | 8 | 5 |
| 18 | 4 | 0 | 0 | 0 | 0 | 5 | 8 | 7 |

**Table 7**. Pure-phase bond energies for dilute $Ni_{1-x}M_xO$ rock salt alloy oxides calculated from a fit of the bond-synergy (BS) model from Eq. (1) to the bond energies shown in Figures 7 and 8 in comparison to a linear model without bond synergy factor.

| Metal-oxygen bond | BS bond energy (eV) | Bond synergy factor (eV) | No-BS bond energy (eV) |
|---|---|---|---|
| Ni-O | -1.4845 | – | -1.4826 |
| Co-O | -1.1365 | -0.2208 | -1.3522 |
| Ru-O | -1.6147 | 0.3013 | -1.3626 |
| Mn-O | -1.8118 | 0.3823 | -1.4872 |
| W-O | -1.8764 | 0.1696 | -1.7425 |
| Fe-O | -1.9038 | 0.4441 | -1.5338 |
| Cr-O | -1.9047 | 0.5222 | -1.4548 |
| Mo-O | -2.0456 | 1.0177 | -1.1519 |

**Table 8.** Bader charges in non-compensating rock salt Ni-based alloy oxides.

| Mo$_x$ | Ni$_{1-x}$ | O |
|---|---|---|
| 1.897-0.903$x$ | 1.292-1.189$x$ | -1.291+0.549$x$ |
| W$_x$ | Ni$_{1-x}$ | O |
| 1.866-1.951$x$ | 1.303-1.425$x$ | -1.303+0.905$x$ |

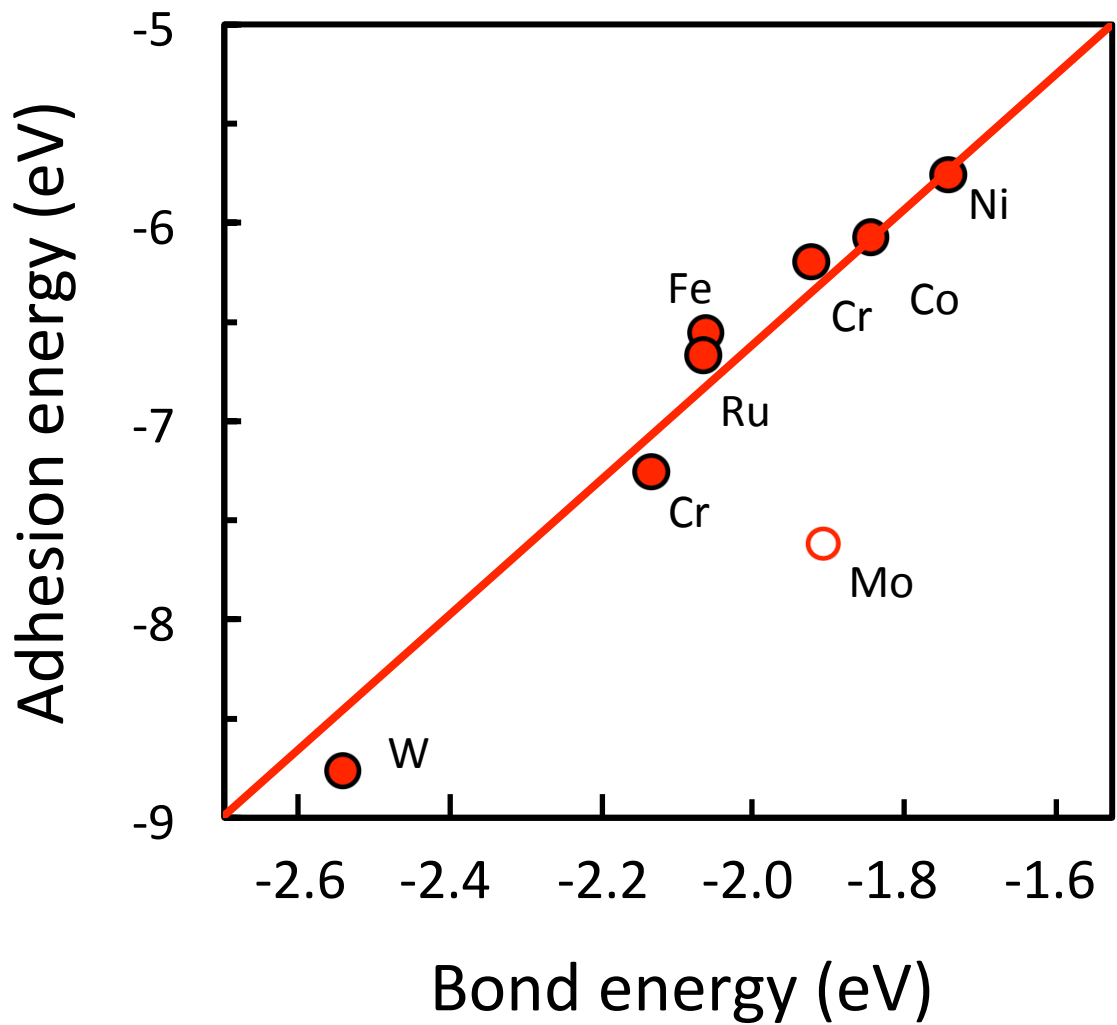

Figure 1. Ground state adhesion energies of O atoms on (001) metal surfaces vs. fitted bond energies $\epsilon_{Mi\text{-}O}$ (Eq. 1), from DFT calculations. The solid circles are results for (in order of increasing energy) W, Cr, Ru, Fe, Mn, Co, and Ni. The hollow circle represents Mo.

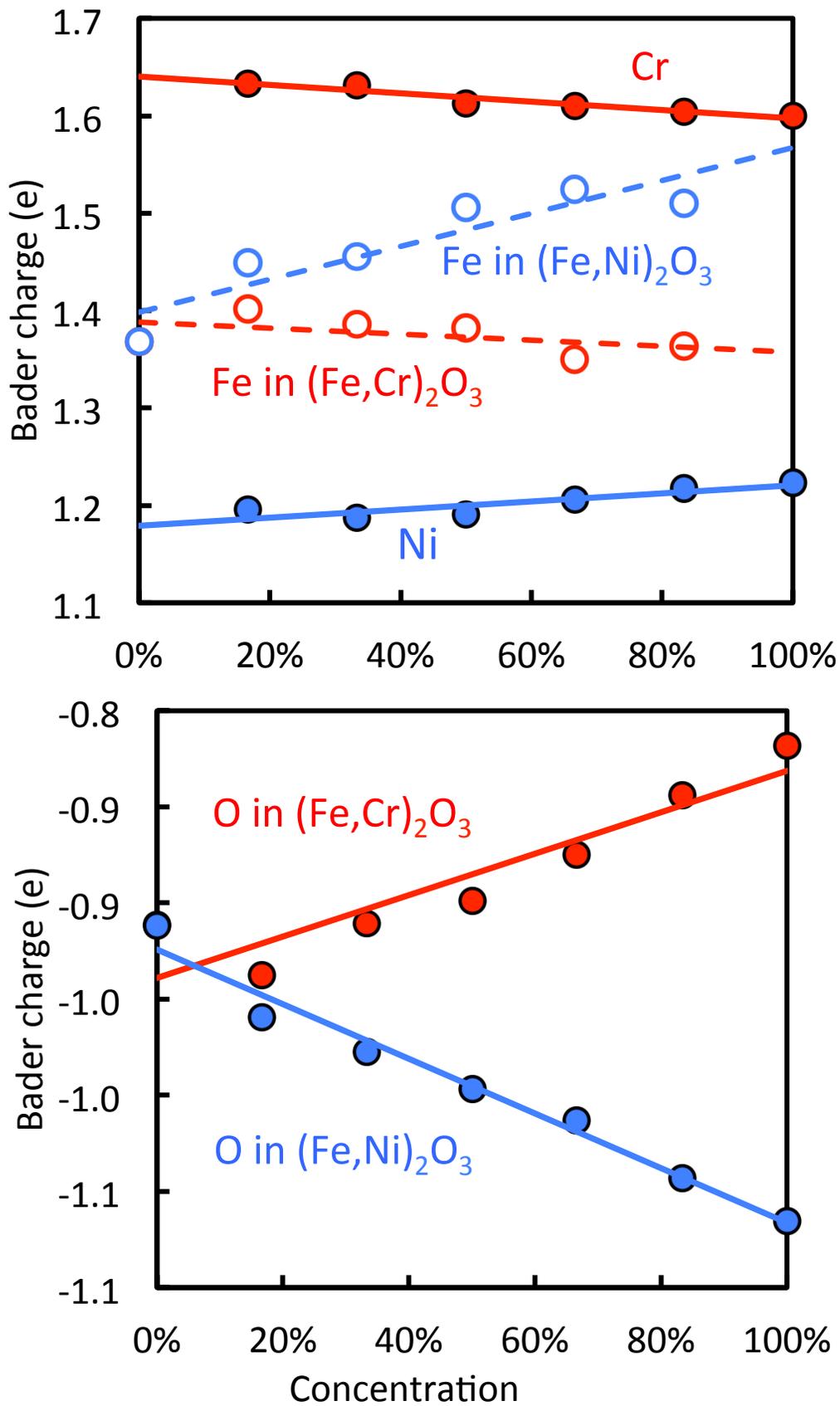

Figure 2. Average Bader charge on (a) cations and (b) oxygen in $(Fe,Cr)_2O_3$ and $(Fe,Ni)_2O_3$ corundum alloys as a function of Ni and Cr solute concentration.

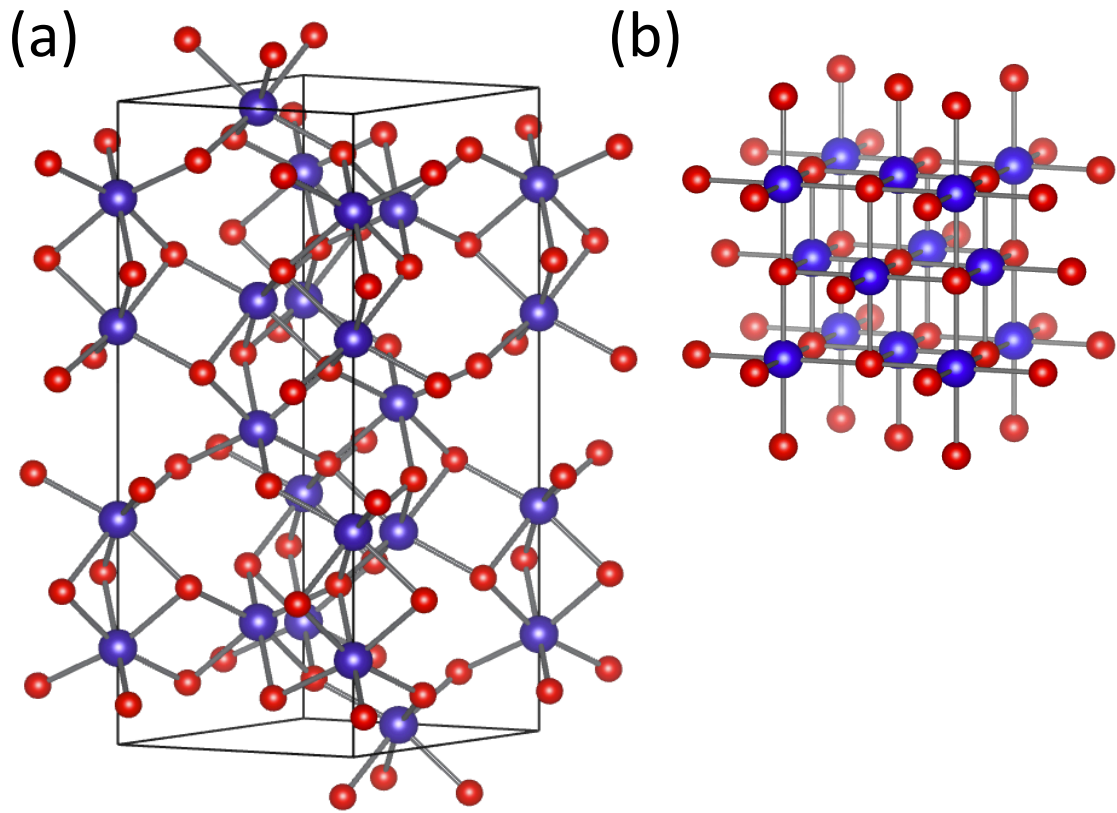

Figure 3. Snapshots of (a) a corundum structure ($M_2O_3$) and (b) a rock salt structure (MO). A blue sphere represents a metal atom and a red sphere represent an oxygen atom.

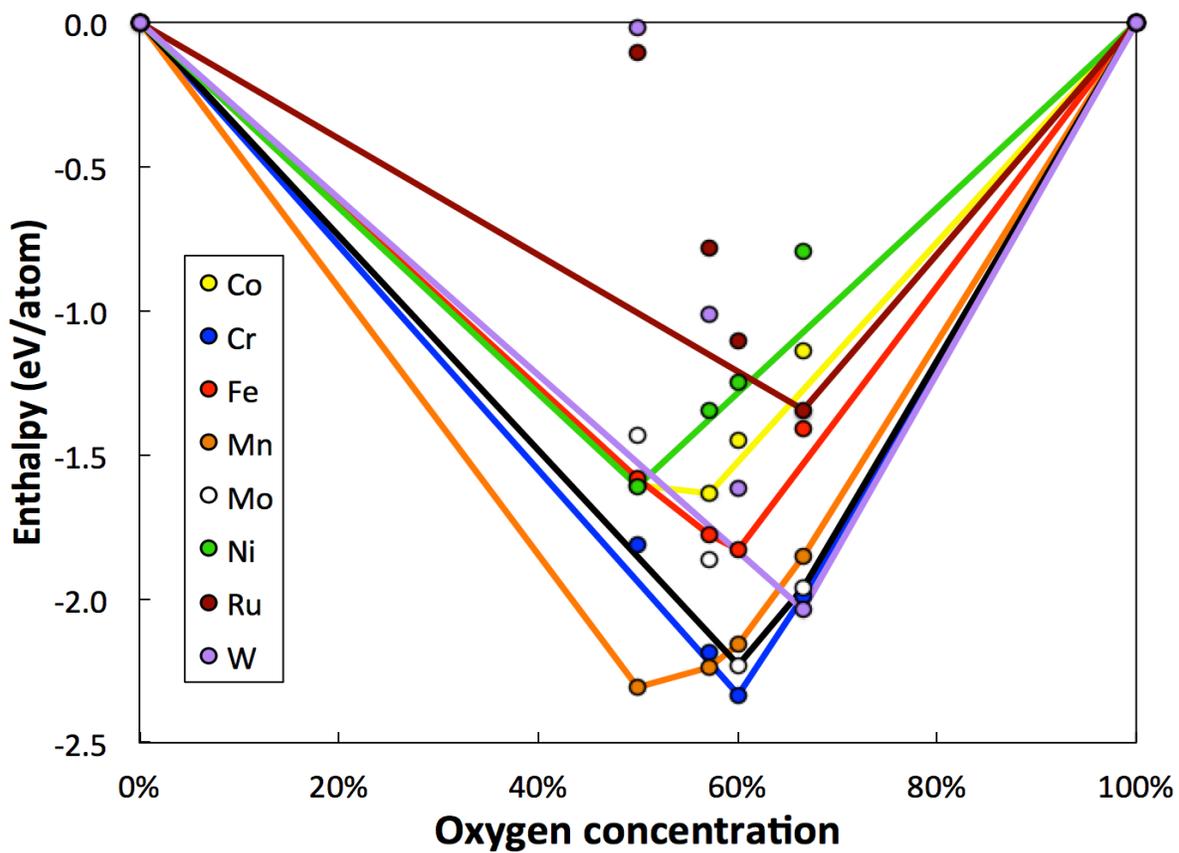

Figure 4. Convex hull of the different oxides as labeled in the legend. The structures considered are rock salt (MO, 50% O), corundum ($M_2O_3$, 60% O), spinel ($M_3O_4$, 57% O) and rutile ($MO_2$, 67% O).

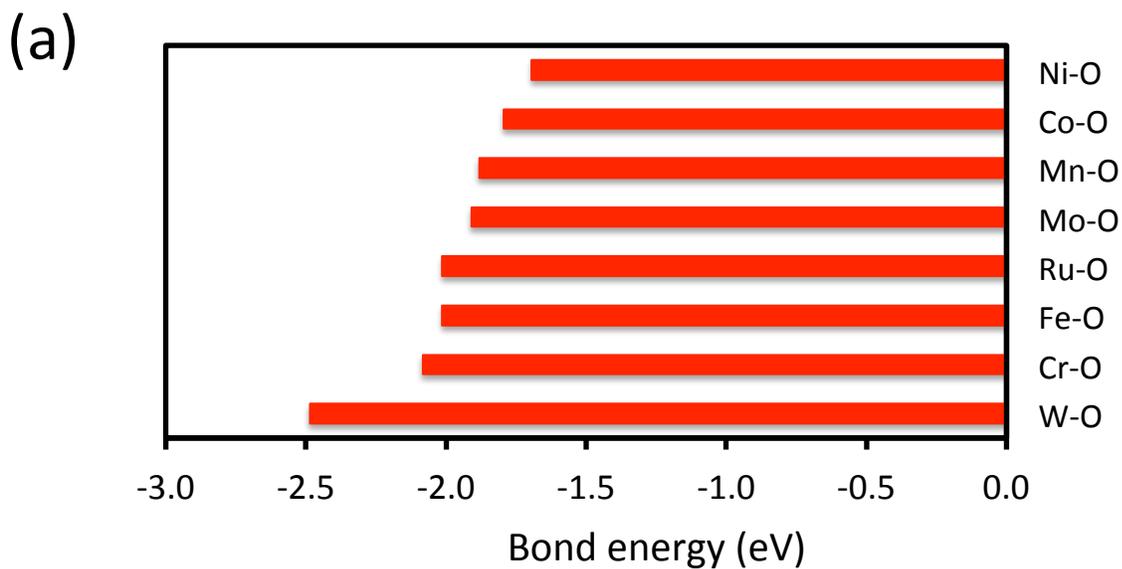

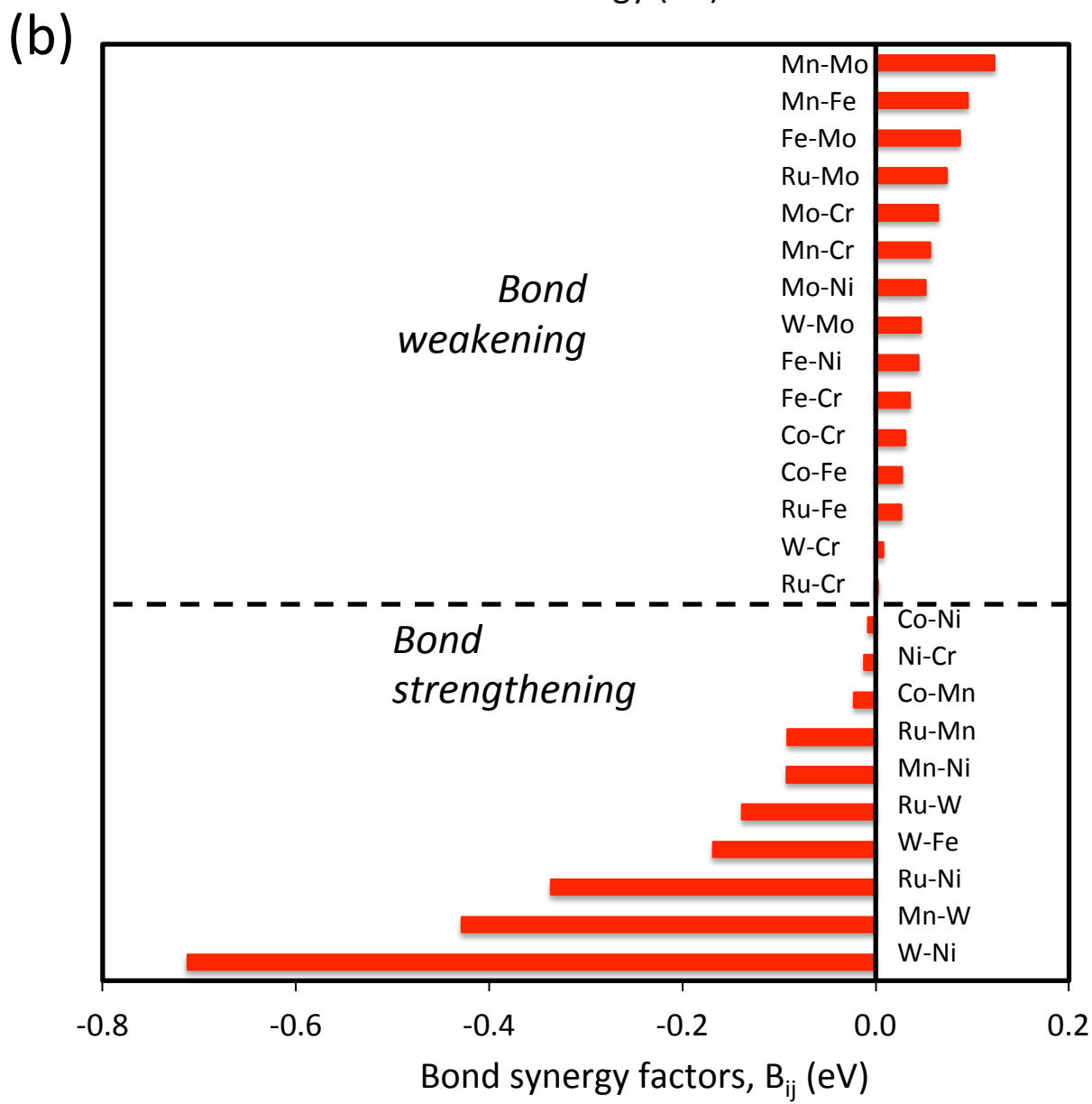

Figure 5. Fitted pure-phase bond energies (a) and bond synergy factors (b) as labeled for the binary corundum alloy oxides shown in Figure S1 in supplementary materials.

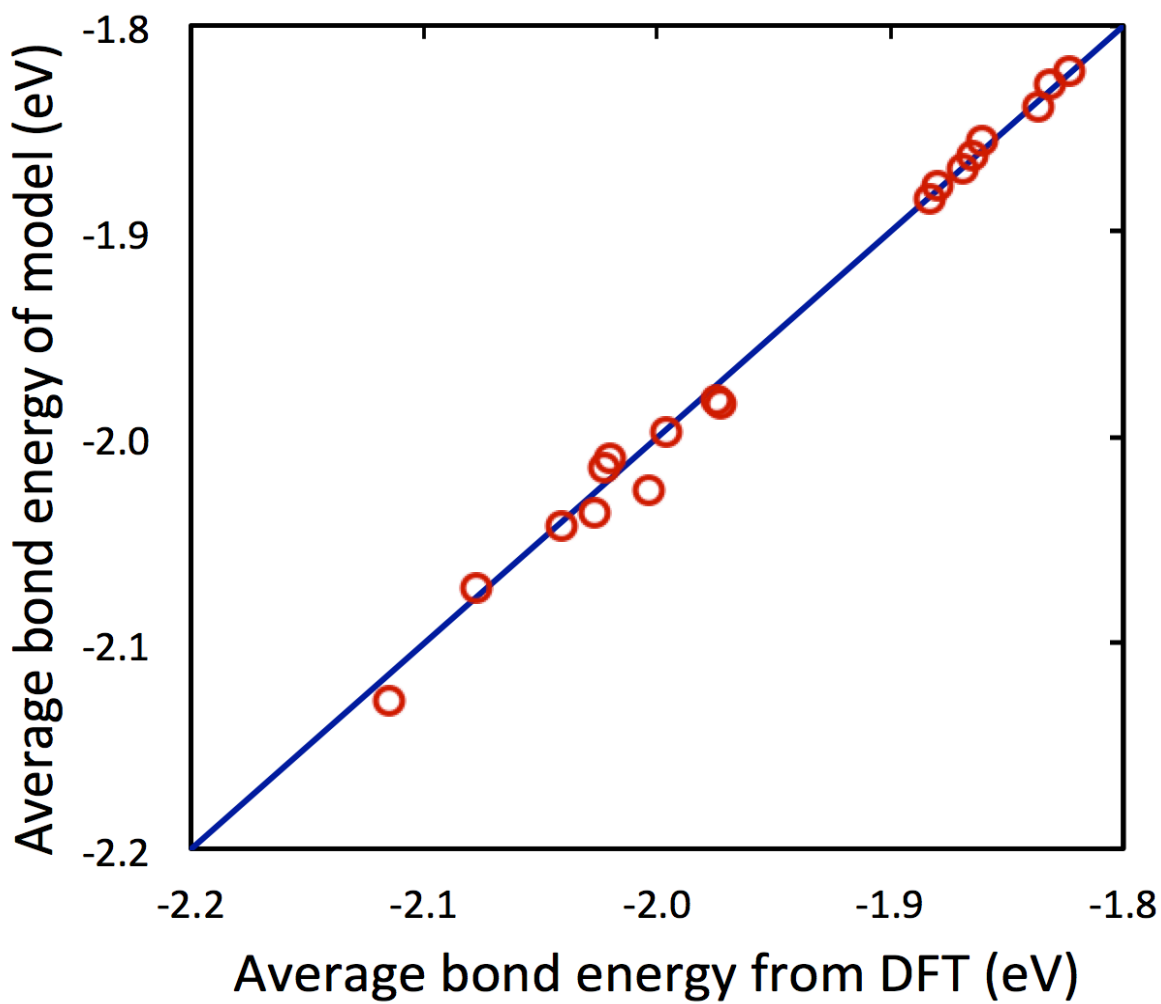

Figure 6. The comparison of average bond energy computed form DFT calculations and from our model.

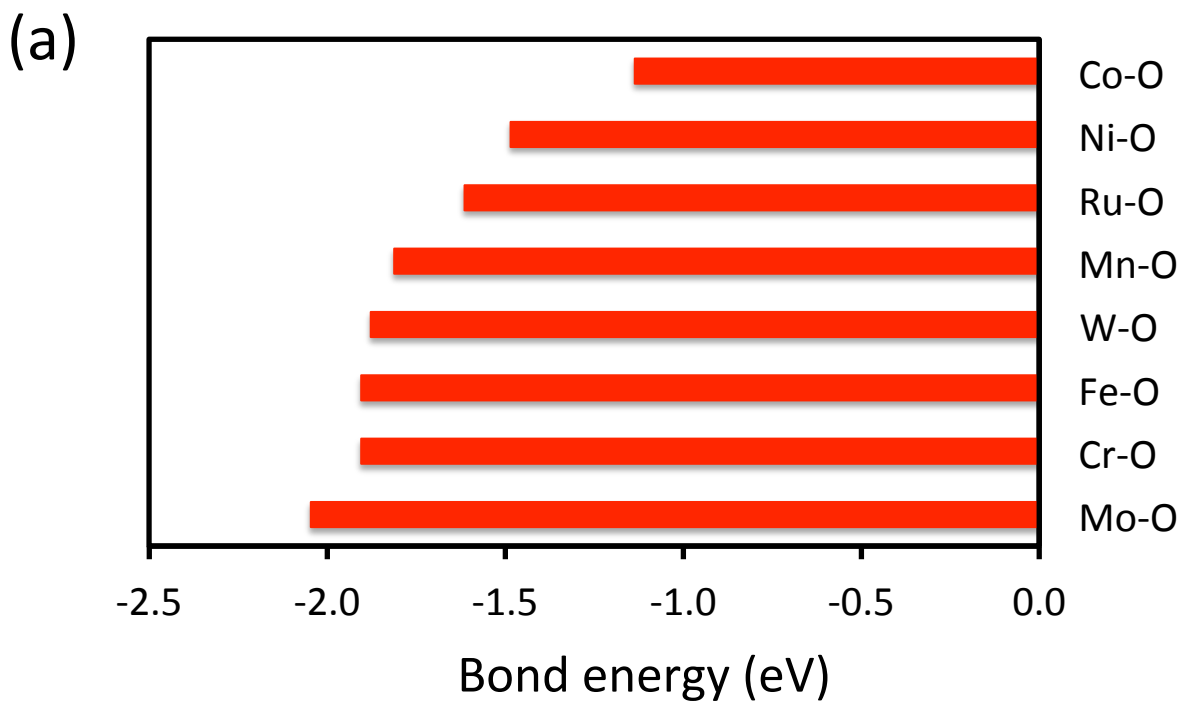
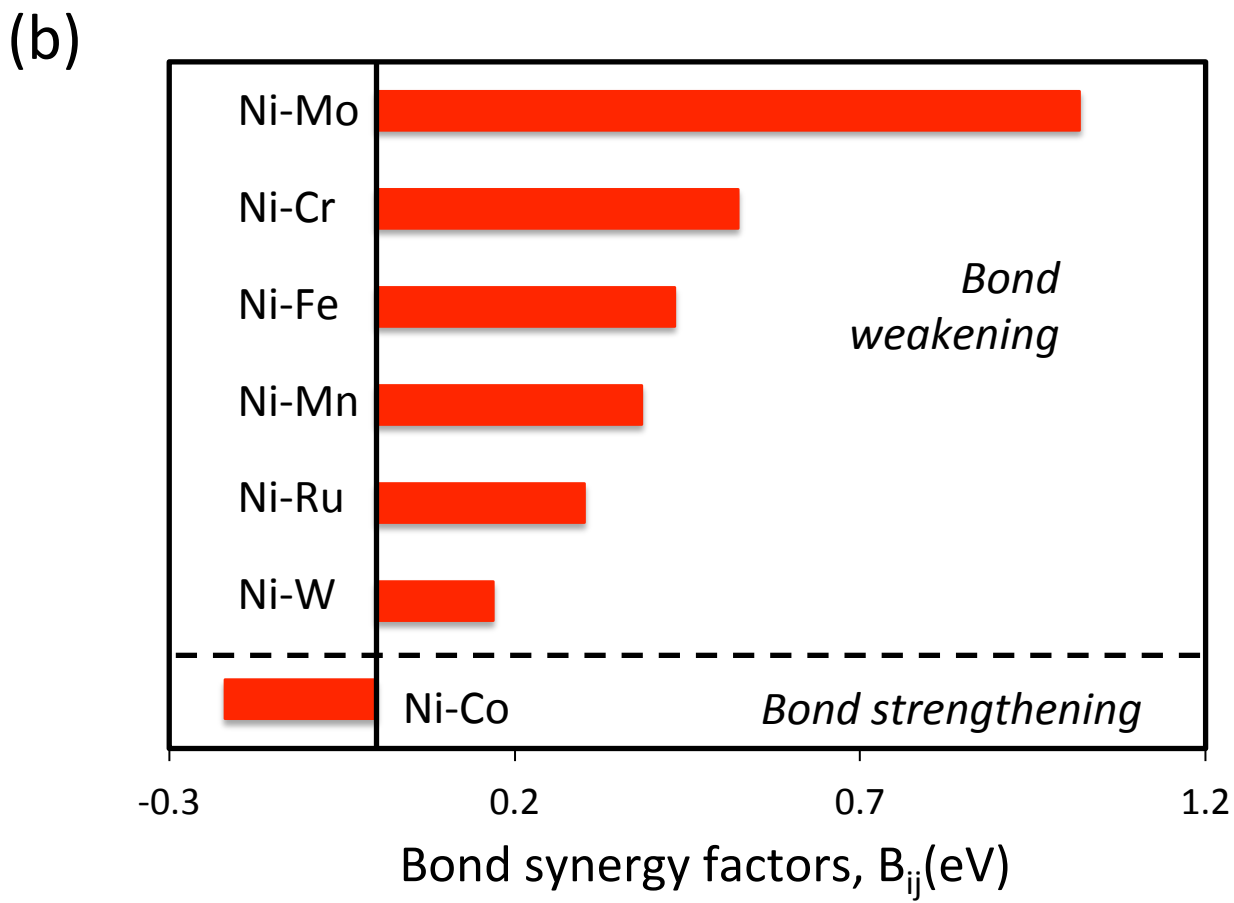

Figure 7. Fitted pure-phase bond energies (a) and bond synergy factors (b) as labeled for the binary rock salt alloy oxides.

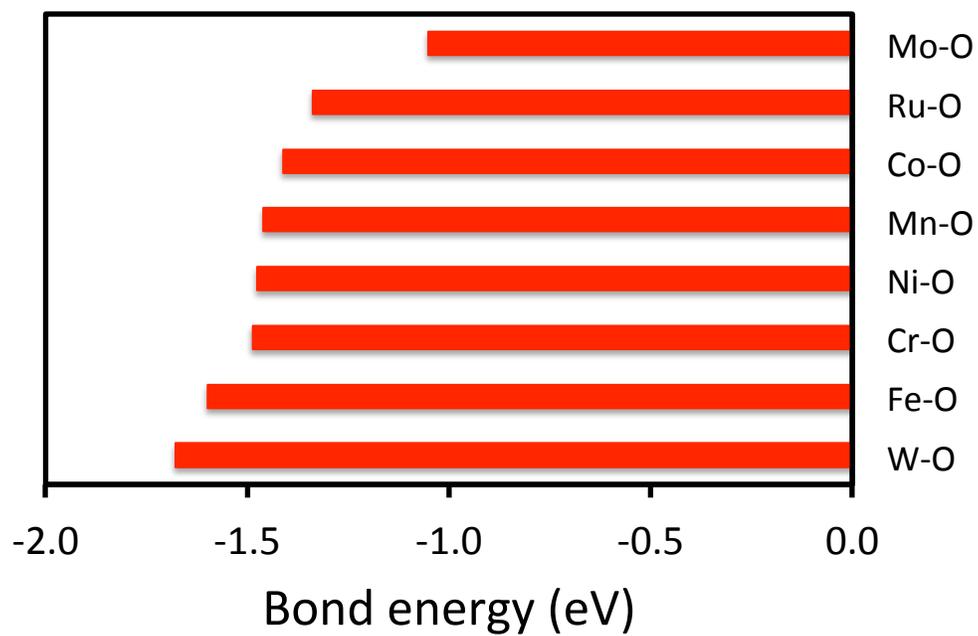

Figure 8. Fitted pure-phase bond energies for the binary rock salt alloy oxides without considering bond synergies.